\documentclass{aa} 
\usepackage{graphicx} 

\def\mkm{{\mu}\rm{m}}

\begin{document}
\title{Reduction of chemical networks}

   \subtitle{I. The case of molecular clouds}

   \author{D. Wiebe\inst{1}, D. Semenov\inst{2},
           \and  Th. Henning\inst{3}
           }

   \offprints{D. Wiebe, \email{dwiebe@inasan.rssi.ru}}

        \institute{Institute of Astronomy of the RAS, 
                   Pyatnitskaya St. 48, 119017 Moscow, Russia\\
                   \email{dwiebe@inasan.rssi.ru}
                \and
                   Astrophysical Institute and University
                   Observatory,
                   Schillerg\"a{\ss}chen 2-3, 07745 Jena, Germany\\
                   \email{dima@astro.uni-jena.de}
                \and
                   Max Planck Institute for Astronomy, 
                   K\"onigstuhl 17, 69117 Heidelberg, Germany\\
                   \email{henning@mpia.de}
                  }
       
   \date{Received ??? / Accepted ???}

\abstract{We present a new method to analyse and reduce chemical
networks and apply this technique to the chemistry in molecular clouds. 
Using the technique, we investigated the possibility of reducing the number
of chemical reactions and species in the UMIST\,95 database
simultaneously. In addition, we did the same reduction but with the 
``objective technique'' in order to compare both methods. We found that 
it is possible to compute the abundance of carbon monoxide and fractional
ionisation accurately with significantly reduced chemical networks in the 
case of pure gas-phase chemistry. For gas-grain chemistry involving surface 
reactions reduction is not worthwhile. Compared to the ``objective  
technique'' our reduction method is more effective but more time-consuming 
as well.
   \keywords{astrochemistry --
             stars: formation --
             molecular processes -- 
             ISM: molecules -- 
             ISM: abundances
            }
        }

\titlerunning{Reduction of chemical networks. I}
\authorrunning{D. Wiebe et al.}

   \maketitle

\section{Introduction}
Despite the significant growth of computer power over the 
last years, the modelling of the chemical evolution of protostellar 
cores and protoplanetary discs is still a challenging 
computational task. This is even more true if one aims at the 
simultaneous self-consistent modelling of the dynamical and chemical 
evolution of these systems.

The mathematical formulation of the chemical processes in space
is realised by differential equations describing chemical kinetics. 
They contain terms for all chemical reactions expected to proceed 
efficiently in the interstellar medium (ISM). Along with abundances 
of all involved species, these terms depend on the rate coefficients 
that are summarised in special databases. The relevant astrochemical
rate files, like the UMIST rate file or Ohio New Standard Database,
comprise hundreds of species and thousands of reactions
(Millar et al.~\cite{UMIST90,UMIST95}; Le Teuff et al.~\cite{UMIST99};
Terzieva \& Herbst~\cite{TH98}; Aikawa \& Herbst~\cite{AH99}).
Of course, one can only wonder if these databases contain all the
information about reactions needed to accurately
reproduce observed abundances of interstellar molecules. The 
opposite question to ask is if all this information is always needed.

In many astronomically interesting situations it is enough to follow
the evolution of a limited number of species only. Intuitively,
it seems to be apparent that to predict an abundance of 
N$_2$H$^+$, for example, one can afford computing C$_{10}$
abundance with less accuracy or even ignore this molecule altogether.
In most cases, though, modelling of the chemical evolution of a parcel 
of the interstellar gas under fixed conditions or in a limited 
predefined range of conditions is not very time-consuming, so one 
probably will not be eager to sacrifice any accuracy even for the sake 
of order of magnitude gain in computational speed. However, in the case 
of dynamical modelling with changing density, UV intensity, temperature 
etc., a reduced chemical network may help to distinguish between 
feasible and non-feasible problems or at least to ease the computational burden
if parallel processors are in use. One possible application for 
reduced chemical networks would be the modelling of the evolution of 
magnetised protostellar clouds or protoplanetary accretion discs, when 
it is necessary to compute the fractional ionisation self-consistently. 
Another example of dynamically important species is carbon monoxide 
(CO) which under certain conditions can be the most important molecular
coolant (see, for example, Glassgold \& Langer~\cite{GL73}).

Several authors have already used reduced chemical networks, 
based on the above arguments, however, without presenting accurate 
mathematical considerations (e.g., Gerola \& Glassgold~\cite{GG80}; Howe, Hartquist, \& Williams
\cite{howe}; El-Nawawy, Howe, \& Millar~\cite{El97}).
Recently, Ruffle et al.~(\cite{Rea02}) (hereafter RRPHH) and Rae et 
al.~(\cite{Rae02}) (hereafter RBHPR) made a first attempt to reduce 
astrochemical networks by so-called ``objective reduction
techniques'', developed in combustion chemistry. They have shown that it 
is possible to quantify the reduction task by introducing a certain 
criterion that would allow in each particular case to indicate, 
abundances of which species must be followed in order to have a 
reasonable estimate for an abundance of a species under investigation.  
For static regions with physical conditions typical of diffuse clouds, 
they isolated the species set that contains tens instead of hundreds 
species. This reduction only introduces a factor of two uncertainty in 
predicted CO abundance. The same task has been solved for the fractional 
ionisation as well.

We perform a similar analysis of the UMIST\,95 chemical reaction database 
in order to find if it is possible to reduce the number of species and/or 
reactions in the chemical networks used to calculate ionisation degree and 
carbon monoxide abundance under conditions typical of molecular clouds. 
Unlike RRPHH and RBHPR, we partly focus on the medium that is more advanced 
toward the formation of a star. This situation is more complicated from
the chemical point of view. First, pre-stellar clouds are often shielded from the 
interstellar UV radiation field. Dissociating photons, being an additional 
chemical factor, in effect simplify the chemical network, as they prevent 
many complex species from being formed. Reactions in grain surfaces are 
another factor that cannot be ignored in dense clouds.

The paper is organised as follows. First, we describe our chemical model
in Sect.~2. The species-based and reaction-based reduction methods are 
outlined in Sect.~3. Results of reductions obtained with both methods are 
presented in Sect.~4 (ionisation degree) and Sect.~5 (CO). These results 
are discussed in Sect.~6. A conclusion follows.

\section{Chemical model}

We take the rates for gas-phase reactions from the UMIST\,95 rate file
(Millar et al.~\cite{UMIST95}). In addition to gas-phase chemistry, 
gas-dust interaction and dust surface reactions are considered in some 
of the models. The rates at which surface reactions proceed
and the very applicability of the rate approximation to surface 
chemistry still remain a subject of debate. This is why many authors
prefer to neglect surface reactions altogether (for instance, see
Aikawa \& Herbst~\cite{AH99}; Bergin \& Langer~\cite{BL97}; Bergin
et al.~\cite{Bea95}; Charnley et al.~\cite{Cea01}). However, in 
the attempt to constrain the chemical network in a cold and dense
molecular cloud core, taking into account gas-dust interactions seems
to be unavoidable, as dust is now widely believed to play a crucial 
role in molecular cloud chemistry, serving both as a sink for 
frozen-out species and as a catalyst for the synthesis of complex 
molecules (see, however, Turner~\cite{T00}). In this paper we seek 
not to reproduce any particular abundance pattern observed in 
molecular clouds, but rather to investigate the very possibility of
reduction when surface reactions are present. This is why we
did not try to incorporate any new ideas about surface chemistry.

\subsection{Physical conditions and initial abundances}

We solve equations of chemical kinetics for the range of fixed physical 
conditions, listed in Table~\ref{physc}, where model designations are 
given as well. Gas and dust temperatures are 10~K in all cases. No direct
UV radiation is present, implying the conditions, typical of obscured 
regions, like cores of molecular clouds, where the visual extinction 
A$_\mathrm{V}\geq10$~mag. Dust grains are assumed to be silicate-like 
spheres with a constant size $a_\mathrm{d}=0.1$~$\mkm$, density 
$\rho_\mathrm{d}=3$~g~cm$^{-3}$, and $10^6$ surface sites for adsorption
(Hasegawa et al. \cite{Hea92}). Dust is supposed to constitute $1\%$ of 
the gas density by mass.

All abundances in the paper are given with respect to the number of
hydrogen nuclei. We use ``high metal'' and ``low metal'' sets of initial 
abundances (e.g. Lee et al.~\cite{Lea98}) quoted in Table~\ref{inabun}. 
In the ``high metal'' set the standard solar elemental composition is 
corrected with a modest depletion of 2 for S and stronger depletions of 
10 for Na, 50 for Si, 60 for Mg, and 110 for Fe. The ``low metal'' values 
contain additional depletion factors of 100 for each of these elements. 
The abundances of all elements but P and Cl are taken from Aikawa et 
al.~(\cite{Aea96}). For P and Cl we take values from Grevesse \& 
Sauval~(\cite{GS98}) and use the same depletion factors as for Fe.
Hydrogen is assumed to be completely in molecular form initially.
All other elements are present in atomic form at $t=0$.

Below details of the adopted model of gas-dust interaction are provided.

\begin{table}
\caption{Model designations.}
\label{physc}
\begin{tabular}{ll}
\hline
Notation & Meaning \\
\hline

DIFF & $n_\mathrm{H}=10^3$~cm$^{-3}$ \\
DENS & $n_\mathrm{H}=10^7$~cm$^{-3}$ \\
GAS  & Gas-phase network ($395\times3864^\mathrm{a}$) \\
GAD  & GAS with accretion and desorption \\
     & ($543\times4593$) \\  
DUST & GAD with surface reactions ($543\times4785$) \\
HM   & ``High metals'' \\
LM   & ``Low metals'' \\

\hline
\end{tabular}
\begin{flushleft}
\footnotesize{$^\mathrm{a}$The network contains 395 species involved 
                          in 3864 reactions.
             }
\end{flushleft}
\end{table}
\begin{table}
\caption{Adopted elemental abundances.}
\label{inabun}
\begin{tabular}{lcc}
\hline
Element & ``High metals'' & ``Low metals''\\
\hline
He   &  $9.75(-2)$ & $9.75(-2)$  \\
C    &  $7.86(-5)$ & $7.86(-5)$  \\
N    &  $2.47(-5)$ & $2.47(-5)$  \\
O    &  $1.80(-4)$ & $1.80(-4)$  \\
S    &  $9.14(-6)$ & $9.14(-8)$  \\
Si   &  $9.74(-7)$ & $9.74(-9)$  \\
Na   &  $2.25(-7)$ & $2.25(-9)$  \\
Mg   &  $1.09(-6)$ & $1.09(-8)$  \\
Fe   &  $2.74(-7)$ & $2.74(-9)$  \\
P    &  $2.16(-8)$ & $2.16(-10)$ \\
Cl   &  $1.00(-7)$ & $1.00(-9)$  \\
\hline
\end{tabular}
\end{table}

\subsection{Gas-grain interactions}

\subsubsection{Neutral accretion}

The accretion rate $k_\mathrm{ac}(i)$ of the $i$th species is given by
\[
k_\mathrm{ac}(i)=\pi a_\mathrm{d}^2v_\mathrm{th}(i)n_\mathrm{d}S,
\]
where $v_\mathrm{th}(i)=\sqrt{{8kT_\mathrm{gas}}/\pi m_i}$ is the thermal 
velocity for the $i$th species, $n_\mathrm{d}$ is the number of dust grains 
per unit volume, $T_\mathrm{gas}$ is the gas temperature, $m_i$ is the 
atomic mass of the $i$th species, and $k$ is the Boltzmann constant. The 
sticking probability $S$ is assumed to be 0.3 for all neutral species except 
for H, He, and H$_2$. The sticking coefficient of atomic hydrogen is 
estimated from Eq.~[3.7] by Hollenbach \& McKee~(\cite{HM79}). Sticking 
probabilities for helium and molecular hydrogen are assumed to be zero.

\subsubsection{Dissociative recombination on grain surfaces}

Following Umebayashi \& Nakano~(\cite{UN80}), we suppose that all grains 
have unit negative charge and thus can act as electron donors in the 
same reactions of dissociative recombination that occur in the gas 
phase. The rate coefficient for reactions of dissociative recombination 
on grain surfaces is given by
\begin{equation}
k_\mathrm{dr}(i)=\sum\limits_j\alpha_j \pi a_\mathrm{d}^2v_\mathrm{th}(i)
                                           n_\mathrm{d}C_\mathrm{ion},
\end{equation}
where 
\begin{equation}
C_\mathrm{ion}=S\left(1+{1.671\cdot10^{-3}\over a_\mathrm{d}T_\mathrm{d}}
\right)
\end{equation}
with $a_\mathrm{d}$ and $T_\mathrm{d}$ being the dust size and temperature,
respectively (Rawlings et al.~\cite{Rea92}). Here the summation is over 
different recombination channels; the quantity $\alpha_j$ denotes the 
probability of a particular channel. Sticking probability $S$ is 0.3 for all 
ions. Products of these reactions are assumed to return to the gas-phase 
immediately.

\subsubsection{Desorption from grain surfaces}

The two desorption processes taken into account in our model are thermal
evaporation and cosmic ray induced desorption.

The thermal evaporation rate $k_\mathrm{th}(i)$ of the $i$th surface species
is
\begin{equation}
  k_\mathrm{th}(i)=\nu_0\exp(-T_\mathrm{D}(i)/T_\mathrm{d}),
\end{equation}
where $T_\mathrm{d}$ is the dust temperature, $kT_\mathrm{D}(i)$ is the 
binding energy for physical adsorption of the $i$th species to the dust 
surface, $\nu_0$ is the characteristic vibrational frequency for the
adsorbed species given (Eq.~[3] from Hasegawa et al.~\cite{Hea92}).

The cosmic ray desorption rate $k_\mathrm{cr}(i)$ of the $i$th surface 
species can be estimated as
\begin{equation}
  k_\mathrm{cr}(i)=3.16\cdot10^{-19}\nu_0\exp(-T_\mathrm{D}(i)/70\,\mathrm{K}).
\end{equation}
The values of $T_\mathrm{D}$ for most species are taken from Hasegawa \& 
Herbst~(\cite{HH93}). There are some species in the UMIST\,95 database 
for which desorption energies are not given in that paper. For these 
species, $T_\mathrm{D}$ is interpolated from values of 
chemically identical species or estimated as
\begin{equation}
  T_\mathrm{D}=50A_i,
\end{equation}
with $A_i$ being the corresponding atomic number, which is a good
approximation for species from Table~4 of Hasegawa \&
Herbst~(\cite{HH93}).

\subsubsection{Surface reactions}

The surface reactions as well as their rates are taken from Hasegawa 
et al.~(\cite{Hea92}) with the appropriate correction for different 
densities. Activation energy of $1000$~K is adopted for O~+~CO surface 
reaction (Hasegawa \& Herbst~\cite{HH93}).

\subsection{Chemical networks}

Three chemical networks are considered. The first one is the pure 
gas-phase network GAS. It consists of electrons, 12~atoms,
137~molecules, and 245~ions (395 species in total) involved in 3864 
gas-phase reactions. The gas-grain network GAD is supplemented by
148 surface species and 729 gas-grain interaction processes
(accretion, desorption, and surface ion recombination). The gas-grain 
network DUST contains 192 additional surface reactions. These networks 
form the basis of our investigations. In the rest of the text we refer to 
them as ``full'' networks. Equations of chemical kinetics are solved with 
the standard DVODE solver. Adopted solver parameters are given in the 
Appendix.

\section{Reduction methods}

To reduce the number of species and/or reactions in a chemical network, one
has to provide a mathematical method designed to estimate the significance 
of particular species or reactions to the evolution of the species under 
consideration. In this paper we elaborate two such methods that are outlined 
in this Section.

\subsection{Species-based reduction}

This technique is derived from the method used by RRPHH and RBHPR. They have 
shown that it is possible to reduce the number of species in a chemical 
network in order to compute abundances of {\em selected\/} species with a 
reasonable accuracy. The species-based reduction technique is based on the 
strategy to specify {\em important\/} species and then to select from the entire 
set only {\em necessary\/} species, that must be included in the chemical 
network in order to compute abundances of important species with a reasonable 
accuracy.

\begin{figure}
\includegraphics[width=0.4\textwidth,clip]{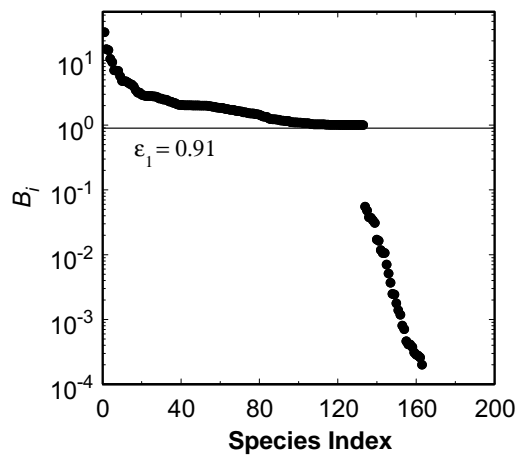}
\caption{Representative $B$ values for the HM-GAS-DIFF model 
computed at $t=10^7$~years. Important species 
are Mg$^+$ and S$^+$. The value of the cut-off 
parameter $\epsilon_1$ is indicated.}
\label{jump1}
\end{figure}

Initially, the reduced species set consists of $N_0$ important species
only. For each species in the entire set we compute the value
\begin{equation}
   B_i=\sum\limits_{j=1,N_0}\left({n_i\over g_j}
       {\partial f_j\over\partial n_i}\right)^2,
\end{equation}
where $n_i$ is the abundance of $i$th species,
\begin{equation}
   f_j=\sum\limits_{lm}k_{lm}n_ln_m-n_j\sum\limits_lk_{jl}n_l=G_j-L_j
\label{f_j}
\end{equation}
is the net rate of the $j$th species abundance change which can be 
expressed as the difference between net gain $G_j$ and loss $L_j$ 
rates, and
\begin{equation}
   g_j=\max(G_j,L_j).
\end{equation}
With these definitions, the quantity
\begin{equation}
   {n_i\over g_j} {\partial f_j\over\partial n_i}
\end{equation}
is the total rate of the abundance change of species $j$ due to those 
reactions that involve species $i$ only. The $B_i$ is a
measure of the sensitivity of the selected species abundances to the 
abundance of the $i$th species. As RRPHH noted,
in this set of $B_i$ values there is a clear boundary between ``needed''
and ``not-so-needed'' species. This is illustrated in Fig.~\ref{jump1},
where representative $B_i$ values and a cut-off threshold $\epsilon_1$ are
shown. We select only those species from the entire set that have $B_i$
above a boundary value, that is
\begin{equation}
B_i\ge\epsilon_1.
\label{bcrit}
\end{equation}
They are added to the reduced set if they are not there already, and $B$
calculation is repeated with the expanded species set. The 
process goes on until at some step no species are added to the reduced 
set. The entire network then consists of these species and of only those
reactions that involve the selected species. Normally, the resultant
reduced set is a combination of several intermediate sets constructed
at different times to account for ``early'' chemistry, ``late'' chemistry etc.
But it turned out that the set reduced at the last time step
($t=10^7$~years) works quite well at earlier times as well.

It should be kept in mind that this method represents only the first
part of the algorithm proposed by RRPHH.

\begin{figure}
\includegraphics[width=0.4\textwidth,clip]{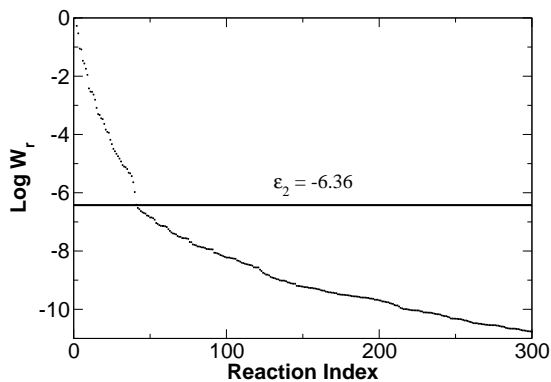}
\caption{Representative $w_\mathrm{r}$ values for the HM-GAS-DENS model 
computed at $t=10^7$~years. An important species 
is CO. The value of the cut-off parameter $\epsilon_2$ is indicated.}
\label{jump2}
\end{figure}

\subsection{Reaction-based reduction}

This new method allows to select only those species and reactions 
from the entire network that are {\em necessary\/} to compute 
abundances of {\em important\/} species with a reasonable accuracy. 
Unlike the species-based reduction technique, where the number of 
reactions is decreased only as a by-product of the lower number 
of species, in this method the numbers of unnecessary species and 
ineffective reactions are reduced {\em simultaneously\/}. In addition 
to the significantly shortened computational time, this could be 
useful if one aims at key destruction and formation 
pathways for important species, where a smaller number of chemical 
reactions makes the analysis easier.

The basic idea of the reaction-based reduction is to search for the 
production and destruction reactions, most important for the evolution of an
important species, and to determine their relative significance. This is done 
via the analysis how sensitive the net formation (or loss) rate of a given
species is to the presence of particular reactions at a certain physical time. 

At the beginning, the chemical model is computed with the full network in 
order to obtain abundances of all species from the network during the entire 
evolutionary time span.

Second, the {\em important\/} species are specified. Further, the algorithm 
estimates the weights of all species $w_\mathrm{s}$ and reactions $w_\mathrm{r}$ 
in order to quantify their importance for the evolution of the species under 
investigation by the following iterative process.

At the first iteration, weights of important species are set to 1 and weights
of all other species are set to 0. At the $i$th iteration all possible formation 
and destruction pathways of the current species $s_i$ are found and their 
significance values $w^i_\mathrm{r}$ are specified as
\[
w^i_\mathrm{r}(j) =
\]
\begin{equation}
\displaystyle \max\left\{w^{i-1}_\mathrm{r}(j),
~\frac{k_jn_{r_1}(j)n_{r_2}(j)}
{\displaystyle \sum_{l=1, N_\mathrm{r}(i)}
k_ln_{r_1}(l)n_{r_2}(l)}w_\mathrm{s}(i)\right\}.
\label{wi_rj}
\end{equation}
Here $k_j$ is the rate of the $j$th reaction, $n_{r_1}(j)$ and 
$n_{r_2}(j)$ are the abundances of the first $r_1(j)$ and second 
$r_2(j)$ reactants in the $j$th reaction, respectively, $N_\mathrm{r}(i)$ is
the number of reactions, in which $s_i$ is a reactant or a 
product, and $w_\mathrm{s}(i)$ is the weight of the species $s_i$.

Consequently, a new set of species, which are found at that 
iteration to be {\em necessary} for the evolution of the important 
species and which were not considered at previous iterations, is 
formed. Their weights $w_\mathrm{s}$ are estimated as the maximum possible
values of the weights $w_\mathrm{r}$ of the reactions they are involved
in. Then these species are added to the set of necessary species,
and the last two steps are repeated. This process is performed for several 
stages during the computational time.

The iterations are stopped when all species and all time moments are 
considered. Then one obtains a reduced chemical network from the 
full network by choosing only those reactions, that have weights $w_\mathrm{r}$ 
exceeding a pre-defined cut-off threshold $\epsilon_2$
\begin{equation}
   \log w_\mathrm{r}\geq \epsilon_2.
\end{equation} 

The typical $w_\mathrm{r}$ values and cut-off parameter $\epsilon_2$ are shown in 
Fig.~\ref{jump2}. The value of this cut-off is chosen to satisfy the requested
accuracy requirement. In our calculations the maximum allowed error is set to 30\%.
If there is a difference greater than 30\% in abundances of the important species 
computed with the full and reduced networks for any particular time step, the 
cut-off is readjusted to a new, smaller value, and the last step is repeated.

\section{Ionisation degree}

We use the two reduction techniques described in the previous section to 
build the reduced chemical networks that could be used for the computation
of the ionisation degree in molecular clouds.  All the models are first 
computed with the full chemical networks. Then we apply the reduction 
techniques and compare the ionisation degree computed with the full and 
reduced networks. The results of the reduction together with relevant 
computational speed gains are summarised in Tables~\ref{diff_e}--\ref{gad_e}. 

\subsection{Diffuse cloud}
\label{gas_diff_e}

\begin{figure}
\begin{center}
\includegraphics[width=0.4\textwidth,clip]{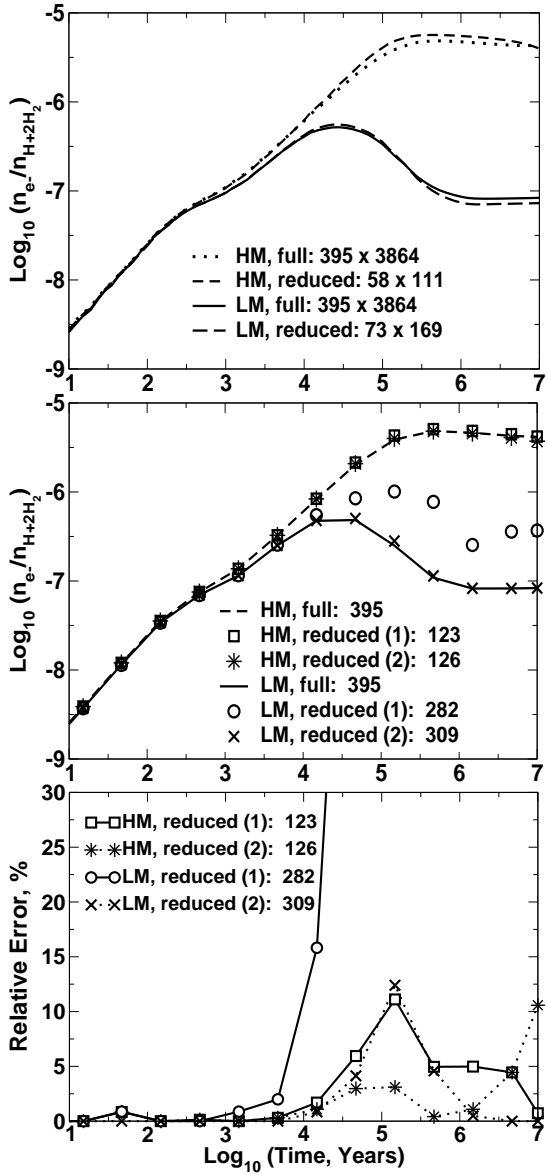}
\end{center}
\caption{Ionisation degree (top and middle panels) and relative 
errors (bottom panel) in a diffuse cloud, computed with the full
and reduced networks for pure gas-phase chemistry (HM-GAS-DIFF and
LM-GAS-DIFF models).
{\em Top} panel reproduces results of the reaction-based
reduction. Indicated are the ``exact'' solutions for high (dotted line) and low
(solid line) metallicities as well as the solution for the reduced species sets
(HM -- dashed line, LM -- long-dashed line).
{\em Middle} and {\em bottom} panels illustrate the results of the 
species-based reduction. Shown are the ``exact'' solutions (HM -- dashed
line, LM -- solid line) and the solutions for various reduced sets (see text).}
\label{diff_dc_e}
\end{figure}

\begin{table}
\caption{Reductions made for ionisation degree in a diffuse cloud.}
\label{diff_e}
\begin{tabular}{lllc}
\hline
Model & Important & Reduced & Speed \\
      & species   & network & gain \\
\hline
\multicolumn{4}{c}{Species-based reduction}\\
\hline
HM-GAS-DIFF & S$^+$         & $123\times1583$    & $\sim25$ \\
HM-GAS-DIFF & S$^+$, Mg$^+$ & $126\times1619$    & $\sim20$ \\
LM-GAS-DIFF & H$^+_3$          & $282\times2539$     & $\sim4$ \\
LM-GAS-DIFF & H$^+_3$, HCO$^+$ & $309\times3177$     & $\sim3$ \\
\hline
\multicolumn{4}{c}{Reaction-based reduction}\\
\hline
HM-GAS-DIFF & e$^-$ & $58\times111$ & 550 \\
LM-GAS-DIFF & e$^-$ & $73\times169$ &  490 \\
\hline
HM-DUST-DIFF & e$^-$ & $210\times683$ & $15$ \\
LM-DUST-DIFF & e$^-$ & $357\times1648$ & $2$ \\
\hline 
\end{tabular}
\end{table}

This case corresponds to the cloud at the initial stage of the protostellar 
collapse ($n_\mathrm{H}=10^3$~cm$^{-3}$). For pure gas-phase chemistry and 
high initial metal abundances (HM-GAS-DIFF), the dominant ions during most 
of the $10^7$~years are S$^+$ and Mg$^+$. We found that it is sufficient to 
consider only these ions as important species in the species-based reduction 
method. Taking sulphur to be the only important species, we can reduce the 
number of species from 395 to 123 and achieve a computational speed gain of a 
factor of about 25 with uncertainty of $\sim15\%$ at $t\sim10^5$~years (see 
Fig.~\ref{diff_dc_e}, middle and bottom panels, squares). If we consider both 
ionised magnesium and ionised sulphur to be important species, the number of 
species in the reduced set is 126, and the accuracy is better than $5\%$ 
during the entire evolution (see Fig.~\ref{diff_dc_e}, middle and bottom panels, 
stars). The computational gain is almost the same (see Table~\ref{diff_e}).
Note that metals other than magnesium (i.e., sodium and iron) are almost 
equally important for the ionisation degree. But as they enter the chemical 
network in a similar way, it is sufficient to designate magnesium to be 
an important species in order to include in the reduced set all the 
species that are needed to compute abundances of Na and Fe as well.

With the reaction-based reduction technique, we choose electrons to be the 
only important species. In this case the number of necessary species is 58,
which is more than two times smaller compared to the species-based reduction.
Additionally, the number of chemical reactions is reduced from 3864 to 111 
(see Fig.~\ref{diff_dc_e}, top panel). The achieved computational time gain is 
of the order of 500, while the resulting uncertainties are by definition smaller 
than $30\%$ during all $10^7$~years of the evolution (see Table~\ref{diff_e}). 
It should be recalled that the accuracy is preset in the reaction-based reduction, 
so in all figures we only show relative errors for the species-based technique.

The small number of necessary reactions in the reduced network can be 
explained by the fact that in this case the chemistry is governed by a rather 
restricted set of ionisation-recombination processes. Such a set is
easily extracted from the entire network by the reaction-based selection 
algorithm. The fundamental reactions during the entire evolution 
time are the ionisation of molecular hydrogen and helium by
cosmic rays and ionisation of carbon and sulphur by cosmic ray induced 
UV photons. These ions rapidly react with abundant molecules by efficient 
ion-molecule reactions, forming complex molecular ions, like HCO$^+$.
In turn, these molecular ions transfer charge to metal atoms. As metal 
ions are less chemically active, they become the most abundant charged 
species, regulating the fractional ionisation (Oppenheimer \& 
Dalgarno~\cite{od1974}).

The situation is more complicated from the chemical point of view when 
the initial metallicity is low. In the absence of abundant metals, the role 
of electron suppliers goes to H$_3^+$, HCO$^+$, and C$^+$, involved in a 
rich chemistry. Even though H$^+_3$ is the dominant ion, it alone does not 
determine the ionisation degree with enough accuracy which is illustrated 
in Fig.~\ref{diff_dc_e} (middle and bottom panels, circles). When this ion 
only is considered to be an important species in the species-based reduction, 
errors are negligible only at earlier times when the chemistry is relatively 
simple. However, after $10^4$~years the errors grow significantly, exceeding 
a factor of 3 at the end of the computation. To account for later chemistry, 
one has to take into account another important ion, HCO$^+$. When these two 
species are designated as important, errors do not exceed $15\%$ during the 
entire computational time (crosses in middle and bottom panels in 
Fig.~\ref{diff_dc_e}). However, due to the complicated chemical connections 
of the two ions, the species set reduction is only modest in this case. Less 
than one hundred species can be excluded from the set with a reduction of 
the computational time by a factor of 3 only (see Table~\ref{diff_e}).
 
Not all the reactions involving these ions are important, though.
With the reaction-based reduction method, we succeed in isolating a three 
time smaller set of 73 species involved in 169 reactions which are 
needed to follow the evolution of the ionisation degree in this 
case. It decreases the time, needed for the computation, by a factor of 
500 (see Table~\ref{diff_e}). The slightly larger number of the necessary 
species and reactions in this case compared to the case of high metal initial 
abundances can be explained by more diverse chemistry involving 
interactions of complex species. The accuracy of the computed
ionisation degree is again better than 30\%.

Inclusion of accretion and desorption processes as well as surface reactions
into these models does not lead to any noticeable changes, primarily, because 
the density is too low to produce significant ice mantles. Results of the 
species-based method show that it is possible to use the same reduced set in 
HM-GAS-DIFF, HM-GAD-DIFF, and HM-DUST-DIFF models without significant loss of 
accuracy. The same is true for the low metallicity models. However, applying 
the reaction-based approach to the gas-grain chemical network including surface 
reactions (DUST-DIFF), we found only moderate reduction as nearly half of the 
species have to be retained in the reduced network. The reason is that the 
surface reactions have very high rates compared to the gas-phase reactions and 
due to the selectivity of the reaction-based method many of them are designated 
to remain in the reduced network. This leads to the larger number of chemical 
species and reactions in such a network compared to the pure gas-phase case 
and, consequently, to the smaller computational time gains, which are about a 
factor of ten in this case (see Table~\ref{diff_e}).

\subsection{Dense cloud}
\subsubsection{Gas-phase chemistry}

When no gas-dust interaction is taken into account, the case of a dense
cloud ($n_\mathrm{H}=10^7$~cm$^{-3}$) at high metallicity is qualitatively
similar to the low-density case. Metals are again the dominant ions.
To compute their abundances (and hence the ionisation degree) with better
than 50\% mean accuracy one needs to include sodium only to the list of
important species in the case of the species-based reduction. The reduced
set contains 72 species and provides two orders of magnitude gain in 
computational speed. To further improve accuracy we add sulphur to the list
of important species. With the reduced set of 121 species we obtain
errors less than 10\% during the entire computational time. The 
reaction-based technique gives similar results and produces a reduced 
network consisting of 131 species and 313 reactions with the same 
computational gain. The difference between the high- and low-metallicity 
cases is smaller for a dense cloud than for a diffuse cloud. In a denser 
environment such chemically active ions as H$^+_3$ and HCO$^+$ are less 
abundant, and metal ions dominate the ionisation degree both in high- and 
low-metallicity cases. As a result of this similarity, in a dense cloud for 
the high and low metallicities the reduced species sets are nearly the same 
both for the species- and reaction-based approaches, with similar errors 
and computational gains (see Table~\ref{dense_e1}).

\begin{table}
\caption{Reductions made for the ionisation degree in a dense cloud with
         pure gas-phase chemistry.}
\label{dense_e1}
\begin{tabular}{lllc}
\hline
Model & Important & Reduced & Speed \\
      & species   & network & gain \\
\hline
\multicolumn{4}{c}{Species-based reduction}\\
\hline
HM-GAS-DENS & Na$^+$        & $72\times568$      & $\sim100$ \\
HM-GAS-DENS & Na$^+$,S$^+$  & $121\times1560$     & $\sim30$ \\
LM-GAS-DENS & Na$^+$        & $71\times567$      & $\sim100$ \\
LM-GAS-DENS & Na$^+$,S$^+$  & $132\times1717$     & $\sim25$ \\
\hline
\multicolumn{4}{c}{Reaction-based reduction}\\
\hline
HM-GAS-DENS & e$^-$ & $131\times313$ & 130 \\
LM-GAS-DENS & e$^-$ & $157\times386$ & 55 \\
\hline
\end{tabular}
\end{table}

\subsection{Gas-grain chemistry}
\label{dust_dens_e}

\begin{figure}
\begin{center}
\includegraphics[width=0.4\textwidth,clip]{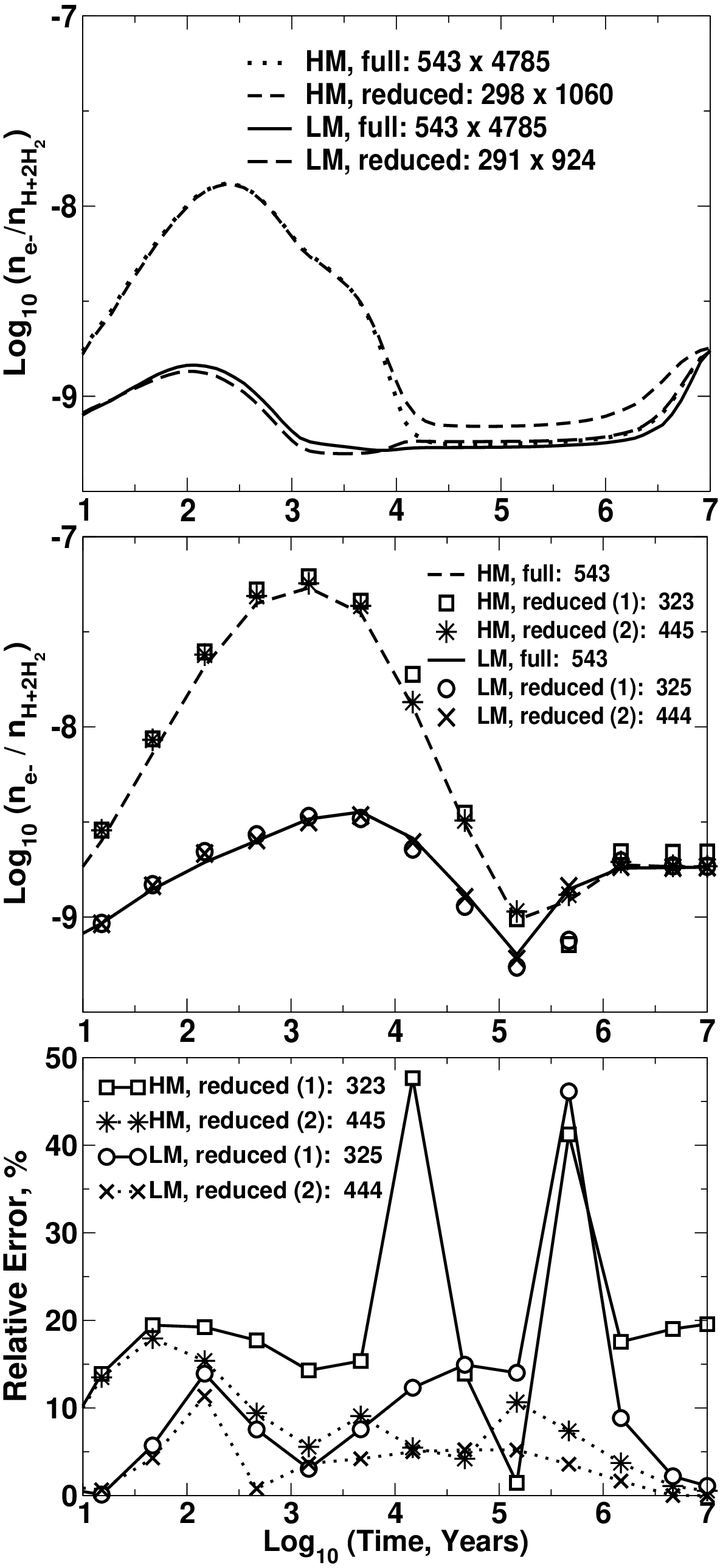}
\end{center}
\caption{Ionisation degree (top and middle panels) and relative
errors (bottom panel) in a dense cloud, computed with the full and
reduced networks for gas-phase chemistry including accretion
and desorption processes and dust surface reactions.
{\em Top} panel reproduces results of the reaction-based reduction 
for the DUST-DENS model.
Indicated are the ``exact'' solutions for high (dotted line) and low
(solid line) metallicities as well as the solution for the reduced 
species sets (HM -- dashed line, LM -- long-dashed line). {\em Middle} 
and {\em bottom} panels illustrate results of the species-based reduction 
for the GAD-DENS model. Shown are the ``exact'' solutions (HM -- dashed 
line, LM -- solid line) and the solutions for various reduced sets 
(see text).}
\label{dens_dd_e}
\end{figure}

The situation changes drastically, when gas-grain interactions are 
taken into account. In the high-metallicity case, at early stages of 
the evolution, the dominant ions are the same as in the diffuse medium,
i.e., metals. However, after $10^4$ years their
depletion becomes important, and the dominant ions are H$_3^+$, 
HCO$^+$, and N$_2$H$^+$. The straightforward use of the species-based
criterion under such conditions leads to the significant loss of accuracy
at this time. The problem is that this criterion is based on the selection 
of species whose {\em abundances} are needed to compute the abundance of an
important species. But the depletion case is not that simple. 
Abundances of mantle species with high desorption energies are definitely 
not important and do not have to be computed with any accuracy. Based on
the formal criterion (\ref{bcrit}), these species are to be excluded from
the reduced set. However, this leads to the exclusion of the 
corresponding accretion reactions from the network, and counterparts of 
the excluded mantle species remain in the gas phase, significantly 
altering its chemistry.

With the species-based reduction method we investigated the GAD-DENS model, 
i.e., the model in which accretion and desorption processes are taken into 
account in addition to gas-phase chemistry. As we have no formal way to 
select species whose {\em absence} in the gas phase is important to compute the 
ionisation degree, we choose to check two artificial possibilities. First, 
we add surface metals to the reduced network as they are primary contributors 
to the electron abundance in the gas phase. Second, we allow all the selected 
neutral species to stick to dust grains, reasoning that if their gas phase 
abundances are deemed important then their depletion is important as well.
Results are shown in Fig.~\ref{dens_dd_e} (middle and bottom
panels). When only metal depletion (with a 
handful of other species selected automatically) is taken into account the 
errors are quite large, sometimes reaching $50\%$. To decrease errors, 
depletion of all selected neutral species is needed. In the latter case, the 
approximate solution differs from the ``exact'' one by as much as $20\%$ but 
later on the difference is much smaller dropping to just a few per cent at 
$t=10^7$~years (see Table~\ref{gad_e}).

The ambiguity of the previous selection procedure is not an issue in the 
reaction-based approach which is specifically designed to find necessary 
formation and destruction pathways rather than necessary species. As it was 
mentioned above, this method of reduction treats gas-grain interactions and 
surface reactions equally with gas-phase reactions. Thus, one does not have to 
worry about which surface species to keep in a reduced set in order to estimate 
the abundances of important species. Therefore, this method works much better in 
the reduction of chemical networks involving complicated processes like gas-grains 
interactions and surface reactions. With the reaction-based technique, we reduced 
the number of species and reactions in the gas-grain chemical network involving
surface reactions (DUST-DENS). We were able to remove about half of the species 
from the full gas-grain network and to reduce the number of reactions by a factor 
of 5 (see Fig.~\ref{dens_dd_e}, top panel). The reduction speeds up the 
computation by a factor of 8 only (see Table~\ref{gad_e}). At times, earlier than 
about $10^4$~years, the ionisation degree is calculated with an accuracy better 
than $10\%$. At $t>10^4$~years, the uncertainties grow but remain less than $30\%$. 
The reason is that it is more difficult to follow the evolution of the ionisation 
degree with the reduced network, since the importance of the complicated gas-grain 
interactions and surface reactions is raising up. In the low-metallicity case 
(LM-GAD-DENS), which is more complex from the chemical point of view, the
reaction-based approach produces a reduced network that consists of 291 species 
and 924 reactions and predicts an ionisation degree with a requested accuracy of 
$30\%$ during the entire evolutionary time. It accelerates the computations by a 
factor of 7 (see Table~\ref{gad_e}).
 
Comparison of top (DUST) and middle (GAD) panels in Fig.~\ref{dens_dd_e} shows 
that there is a difference between the evolution of ionisation degree in the case 
of gas-grain chemistry with and without surface reactions. When surface 
reactions are accounted for, fractional ionisation drops to $10^{-9}$ in 
$t\sim 10^4$~years. In the case of gas-phase chemistry with accretion and 
desorption it reaches the same equilibrium value later, after $10^5$~years. Also, 
the maximum value of the ionisation degree is higher in the latter case. Thus, we 
conclude that surface reactions are important for the accurate estimation of the 
fractional ionisation in dense molecular clouds at late stages of their evolution.

To conclude this section, we emphasise that the reaction-based technique is more 
suitable for the reduction of gas-phase or gas-grain chemical networks if one 
intends to follow the evolution of the ionisation degree in diffuse or dense 
molecular clouds with a relatively small number of chemical species and reactions 
and a reasonable accuracy than it is the case for the species-based scheme.

\begin{table}
\caption{Reductions made for the ionisation degree in a dense cloud with
gas-grain chemistry included.}
\label{gad_e}
\begin{tabular}{lllc}
\hline
Model & Important & Reduced  & Speed \\
      & species   & network & gain \\
\hline
\multicolumn{4}{c}{Species-based reduction}\\
\hline
HM-GAD-DENS & H$^+_3$,   & $323\times3201$   & $\sim6$ \\
        & surface metals    &         &         \\
HM-GAD-DENS & H$^+_3$,   & $445\times3445$   & $\sim2$ \\
        & all surface       &         &         \\ 
\hline
LM-GAD-DENS & H$^+_3$,          & $325\times2933$   & $\sim4$ \\
        & surface metals   &         &     \\
LM-GAD-DENS & H$^+_3$,          & $444\times3171$   & $\sim2.5$ \\
        & all surface      &         &     \\ 
\hline
\multicolumn{4}{c}{Reaction-based reduction}\\
\hline
HM-DUST-DENS & e$^-$ & $298\times1060$ & $8$ \\
LM-DUST-DENS & e$^-$ & $291\times924$ & $7$ \\
\hline 
\end{tabular}
\end{table}

\section{Carbon monoxide}

Essentially the same scheme of reduction, as in the previous section, 
is utilised for CO as well. First, the chemical model is computed with 
one of the full chemical networks. Its results are used to select 
key species needed to predict the CO abundance. Then the validity of a 
reduced network is checked by comparing the abundances obtained with the 
full and reduced networks for the entire evolutionary time. In all cases, 
we assumed that the CO molecule is the only important species. Results of 
the reduction and relevant computational speed gains are presented in 
Tables~\ref{diff_co}--\ref{dens_co2}. 

\subsection{Diffuse cloud}
\label{gas_diff_co}

\begin{figure}
\begin{center}
\includegraphics[width=0.4\textwidth,clip]{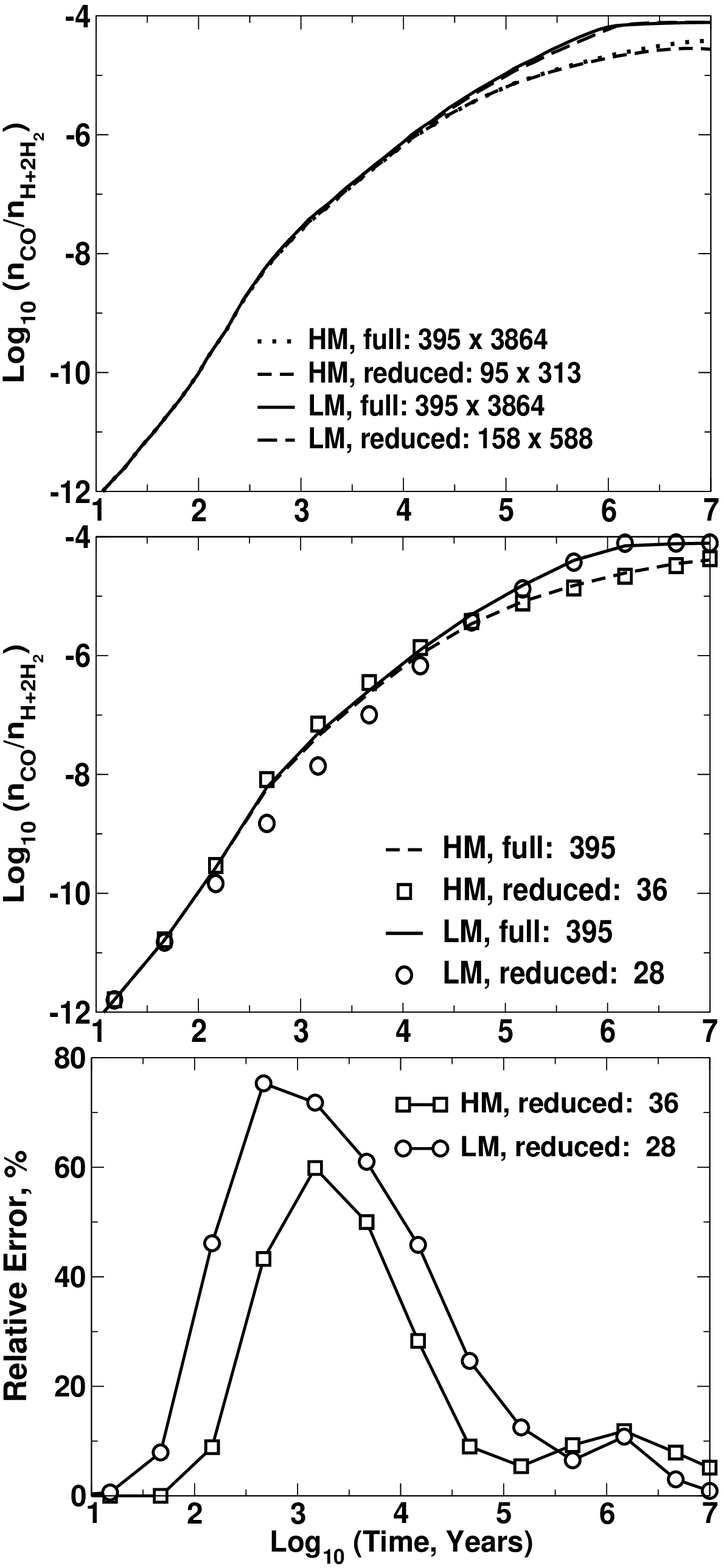}
\end{center}
\caption{The CO abundances (top and middle panels) and relative 
errors (bottom panel) in a diffuse cloud, computed with the full 
and reduced species set for pure gas-phase chemistry (GAS-DIFF model). 
{\em Top} panel reproduces results of the reaction-based
reduction. Indicated are the ``exact'' solutions (HM -- dotted line, LM --
solid line) and the solution for the case when CO is considered to be 
the only important species (HM -- dashed line, LM -- long-dashed line).
The {\em middle} and {\em bottom} panels illustrate results of the species-based 
reduction. Shown are the ``exact'' solutions (HM -- dashed 
line, LM -- solid line) and the solution for the case when CO is 
an important species (HM -- open squares, LM -- open 
circles).}
\label{diff_dc_co}
\end{figure}

In the case of pure gas-phase chemistry in a diffuse cloud (at 
the gas density $n_\mathrm{H}=10^3$~cm$^{-3}$, GAS-DIFF) the main processes, 
relevant to the evolution of carbon monoxide, are its destruction by cosmic 
rays or by cosmic ray induced photons and its production through the 
neutral-neutral reactions of oxygen with carbon atoms and various light 
carbon-bearing molecules, dissociative recombination of HCO$^+$, and 
destruction of H$_2$CO by CR-induced photons. Since metals as well as S, Si, P, and 
Cl do not affect the evolution of CO abundances directly, it is natural to 
expect that there should not be large differences between the high- and low- 
metallicity cases.

In the absence of gas-grain interactions the evolution of CO is
comparatively simple -- during almost the entire evolutionary time its abundance 
is steadily approaching the equilibrium value of about $10^{-4}$ which
is reached at $\sim 10^6$~years (see Fig.~\ref{diff_dc_co},
top and middle panels).

With the species-based reduction, of all 395 species we select only
36 species in the high-metallicity case and 27 species in the case of low metal 
abundances (see Fig.~\ref{diff_dc_co}, middle panel). Due to the small size of 
these reduced networks, time needed to solve the relevant ODE system is negligible,
less than a few seconds (see Table~\ref{diff_co}). At early evolutionary phase 
($t\sim10^3-10^4$~years) the uncertainties of the CO abundance can reach 50\% for 
high metallicity and 80\% for low metallicity. Later, when the equilibrium abundance 
is reached, error drops to about 20\% (see Fig.~\ref{diff_dc_co}, bottom panel).
The reason for the high uncertainty at early times is that such rare species as
CH$_3$CO$^+$ and CH$_2$CO, which are not recognised by the species-based algorithm 
to be necessary species, are in fact important at the early evolutionary phase for 
the accurate estimation of the CO abundance.

Applying the reaction-based reduction, we are able to follow the evolution of
the CO abundance for $10^7$~years with an accuracy better than $30\%$.
However, the price to pay for this is a significant number of 
species to be held in the corresponding reduced networks. One needs to keep 158 
and 95 species in the reduced networks for high and low metallicities,
respectively (see Fig.~\ref{diff_dc_co}, top panel). The relevant computational 
speed gain is a factor of 50 for the HM case and about 200 for the LM case (see 
Table~\ref{diff_co}). It can be seen from Fig.~\ref{diff_dc_co} (top and middle 
panels), that the CO abundances in the case of high- and low-metallicities do behave 
similarly. Nevertheless, from the chemical point of view the case of low initial 
abundances of metals is somewhat simpler in comparison with the high metallicity case 
and implies smaller number of species and reactions to be kept in the reduced network.

Inclusion of accretion-desorption processes and surface reactions does not
affect the chemistry in a diffuse cloud significantly. As we already mentioned in
Sect.~\ref{gas_diff_e}, the gas-grain interactions are not efficient at such low 
density ($n_\mathrm{H}=10^3$~cm$^{-3}$) since the typical rate of collisions 
between dust grains and gas species is far too low. Using the species-based approach, 
we found that it is enough to retain only 37 species for the high metal initial 
abundances and 27 species in the LM case (see Table~\ref{diff_co}). These are almost 
the same numbers as in the case of pure gas-phase chemistry. In fact, one
can use the same reduced sets of species and reactions in the GAS-DIFF, GAD-DIFF,
and DUST-DIFF models (separately for each metallicity set). However, uncertainties
in the computed CO abundances in all cases exceed 50\% at $t\sim10^3-10^4$ years.

It is not that easy to remove these uncertainties. The application of the
reaction-based method allows to cut the entire 543 species set by a factor of 2
in this case (see Table~\ref{diff_co}). It is more than two times the number of 
species found to be necessary for the correct reproduction of CO abundances in 
the case of pure gas-phase chemistry. Consequently, it speeds up the 
calculations by a factor of 5 only. The reason is, as noted in the previous section, 
that the reaction-based algorithm tends to keep some surface reactions in the 
reduced networks.
Obviously, it leads to the larger number of species to be retained in the relevant 
reduced networks compared to the case of pure gas-phase chemistry.

\begin{table}
\caption{Reductions made for CO in a diffuse cloud.}
\label{diff_co}
\begin{tabular}{lllc}
\hline
Model & Important & Reduced & Speed \\
      & species   & network & gain \\
\hline
\multicolumn{4}{c}{Species-based reduction}\\
\hline
HM-GAS-DIFF & CO            & $36\times240$     & $>100$ \\
LM-GAS-DIFF & CO            & $28\times182$     & $>100$ \\
\hline
HM-GAD-DIFF & CO            & $37\times242$     & $>100$ \\
LM-GAD-DIFF & CO            & $27\times173$     & $>100$ \\
\hline
\multicolumn{4}{c}{Reaction-based reduction}\\
\hline
HM-GAS-DIFF & CO & $158\times588$ & 50 \\
LM-GAS-DIFF & CO & $95\times313$  & 240 \\
\hline 
HM-DUST-DIFF & CO & $298\times1293$ & 8 \\
LM-DUST-DIFF & CO & $306\times1310$ & 5 \\
\hline
\end{tabular}
\end{table}

\subsection{Dense cloud}
\subsubsection{Gas-phase chemistry}

\begin{figure}
\begin{center}
\includegraphics[width=0.4\textwidth,clip]{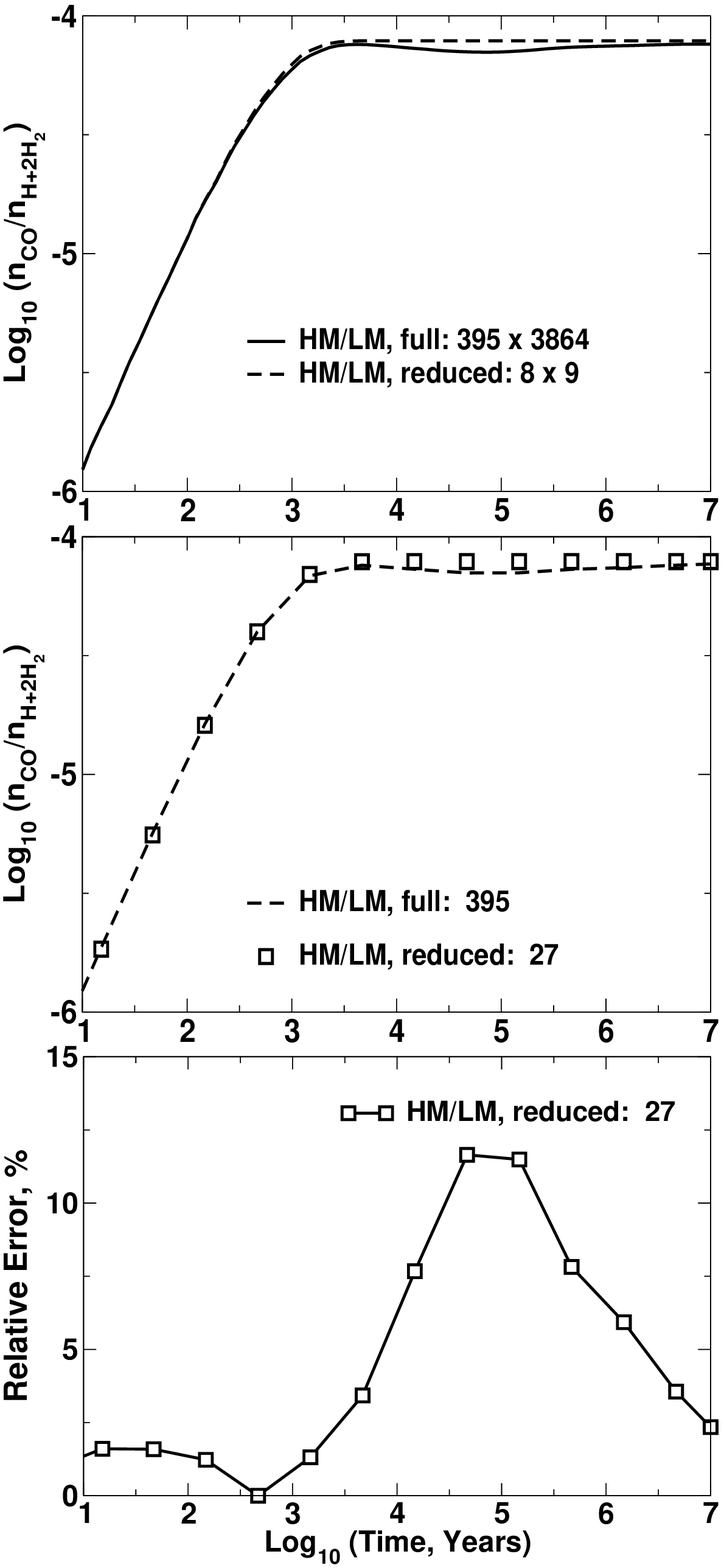}
\end{center}
\caption{Same as in Fig.~\ref{diff_dc_co} but for a dense cloud (GAS-DENS).}
\label{dens_dc_co}
\end{figure}

In a dense molecular cloud ($n_\mathrm{H}=10^7$~cm$^{-3}$) the main processes 
leading to the production and destruction of CO molecules in the case of pure 
gas-phase chemistry (GAS-DENS) are the same as in a diffuse cloud, but with 
one important change. At such high density, the role of chemical reactions
involving abundant species like C, O, CH, and CH$_2$, becomes even more important. 
Therefore, we may expect that in this case the chemistry relevant to the evolution 
of carbon monoxide is even simpler than in the diffuse cloud (GAS-DIFF). It 
implies a smaller set of necessary species.

The results of reduction for this case are shown in Fig.~\ref{dens_dc_co}. 
The CO abundance increases rapidly at early evolutionary 
stages ($t\la 10^3$~years). Later, when the chemical equilibrium is reached,
it remains constant around $10^{-4}$.

Using the species-based reduction technique, we succeeded in the isolation of a
small set of 27 species allowing us to follow the evolution of CO with a very
good accuracy $<15\%$ for all $10^7$~years for both the high- and low-metallicity 
case (see Fig.~\ref{dens_dc_co}, middle and bottom panels). Again, the computational 
time is negligibly small, less than a second (see Table~\ref{dens_co1}).

But even this 27 species set contains excessive information.
Applying the reaction-based reduction approach, we revealed that in reality
there are only 8 species involved in 9 reactions which are necessary for
the accurate estimation of CO abundances during the entire evolutionary time 
(see Fig.~\ref{dens_dc_co}, top panel). The relevant reduced network is given
in Table~\ref{CO}. The uncertainties of computed CO abundances do not exceed 
$15\%$ for all $10^7$~years of the evolution.
This remarkable result, showing the power of the reaction-based technique, 
explains why there is no difference between the cases of high and low metals in a
dense cloud: metals are completely excluded from this network.

In the medium with the atomic initial composition (except for molecular hydrogen)
reactions start from the dissociation of H$_2$ molecules by cosmic ray particles
and from their radiative association with atomic carbon. The latter reaction leads to
CH$_2$ formation. This molecule reacts with atomic H and C leading to
CH and C$_2$H. These three molecules (CH, CH$_2$, and C$_2$H) associate with atomic
oxygen finally producing the CO molecule. These processes are balanced by destruction
of CO and CH by CR-induced photons. In this way, the key reaction pathways can be 
isolated for which high-quality experimental reaction rate measurements have to be
provided for an accurate modelling. It may seem to be somewhat surprising that
the reduced network does not contain HCO$^+$, at variance with the diffuse cloud case.
Note, however, that the maximum value of the HCO$^+$ abundance reached in the
high-density model ($\sim10^{-10}$) is two orders of magnitude less than in
the diffuse cloud model. Given the correspondingly low electron abundance, this prevents
effective CO synthesis in dissociative recombination of HCO$^+$. Proton transfer reactions
of CO with H$_2^+$ and H$_3^+$ that would transform CO to HCO$^+$ are also nearly
quenched due to low abundances of these ions. Thus, while CO is definitely an
important species for HCO$^+$ in a dense medium, the opposite is not true.

Going somewhat out of the scope of the present paper, we mention briefly that this
network proved to be valid under a much wider range of conditions that are described
here. Calculations that will be presented in a subsequent paper show that it can be
used for estimating the CO abundance not only in dark dense clouds, but in a
translucent ($A_\mathrm{V}>1$ mag) medium as well. It gives less than 50\% error at
$10^4<n_\mathrm{H}<10^{10}$~cm$^{-3}$ and $T<250$~K and can be used in a denser medium 
as well ($n_\mathrm{H}<10^{12}$~cm$^{-3}$) provided $T\sim200-250$~K and 
$A_\mathrm{V}>5$ mag, conditions being typical of hot cores or protoplanetary discs.

The accuracy of this network is also insensitive to the variation of the adopted C/O
ratio (0.43) within a factor of a few. The value of this ratio
would be critical, though, in the attempt to construct a reduced network for carbon
or oxygen separately. If carbon is less abundant than oxygen (as in the adopted
abundance sets), all C atoms are locked in CO molecules, and the reduction is rather
effective. In fact, it is possible to construct a single network of a few tens of
species that reproduces abundances of C, O, and CO simultaneously with less than
100\% uncertainty. If the C/O ratio is greater than 1, extra carbon is available
for producing long carbon chains, and the effective reduction (with C as an important species) is
not possible.

\begin{table}
\caption{Reductions made for CO in a dense cloud with pure gas-phase chemistry.}
\label{dens_co1}
\begin{tabular}{lllc}
\hline
Model & Important & Reduced & Speed \\
      & species   & network & gain \\
\hline
\multicolumn{4}{c}{Species-based reduction}\\
\hline
HM-GAS-DENS & CO            & $27\times133$     & $>100$ \\
LM-GAS-DENS & CO            & $27\times152$     & $>100$ \\
\hline
\multicolumn{4}{c}{Reaction-based reduction}\\
\hline
HM-GAS-DENS & CO & $8\times9$ & $> 10^4$ \\
LM-GAS-DENS & CO & $8\times9$ & $> 10^4$ \\ 
\hline
\end{tabular}
\end{table}

\begin{table}
\caption{The reduced gas-phase network for CO chemistry in a dense cloud
obtained with the reaction-based method.}
\label{CO}
\begin{tabular}{lll}
\hline
&Reactions &\\
\hline
    1 H + CH$_2$ &$\rightarrow$& CH + H$_2$ \\
    2 H$_2$ + C &$\rightarrow$& CH$_2$ + PHOTON \\
    3 C + CH$_2$ &$\rightarrow$& C$_2$H + H \\    
    4 O + CH &$\rightarrow$& CO + H \\    
    5 O + CH$_2$ &$\rightarrow$& CO + H + H \\
    6 O + C$_2$H &$\rightarrow$& CO + CH \\      
    7 H$_2$ + CRP$^\mathrm{a}$ &$\rightarrow$& H + H \\       
    8 CH + CRPHOT$^\mathrm{b}$ &$\rightarrow$& C + H \\       
    9 CO + CRPHOT &$\rightarrow$& C + O \\  
\hline
\end{tabular}
\begin{flushleft}
\footnotesize{$^\mathrm{a}$Cosmic ray particle\\
              $^\mathrm{b}$CRP-induced photon}
\end{flushleft}
\end{table}

\subsubsection{Gas-grain chemistry}

The situation is no longer trivial when gas-grain interactions and
surface reactions are taken into account (GAD-DENS and DUST-DENS).
The influence of these processes on the evolution of
abundances is twofold. First, the freeze-out of the gas-phase CO
leads to the activation of certain chemical reaction chains that
are suppressed in the presence of CO. Second, the presence of CO in
icy mantles can drive some surface reactions that may eventually result
in the formation of complex organic species.

\begin{figure}
\begin{center}
\includegraphics[width=0.4\textwidth,clip]{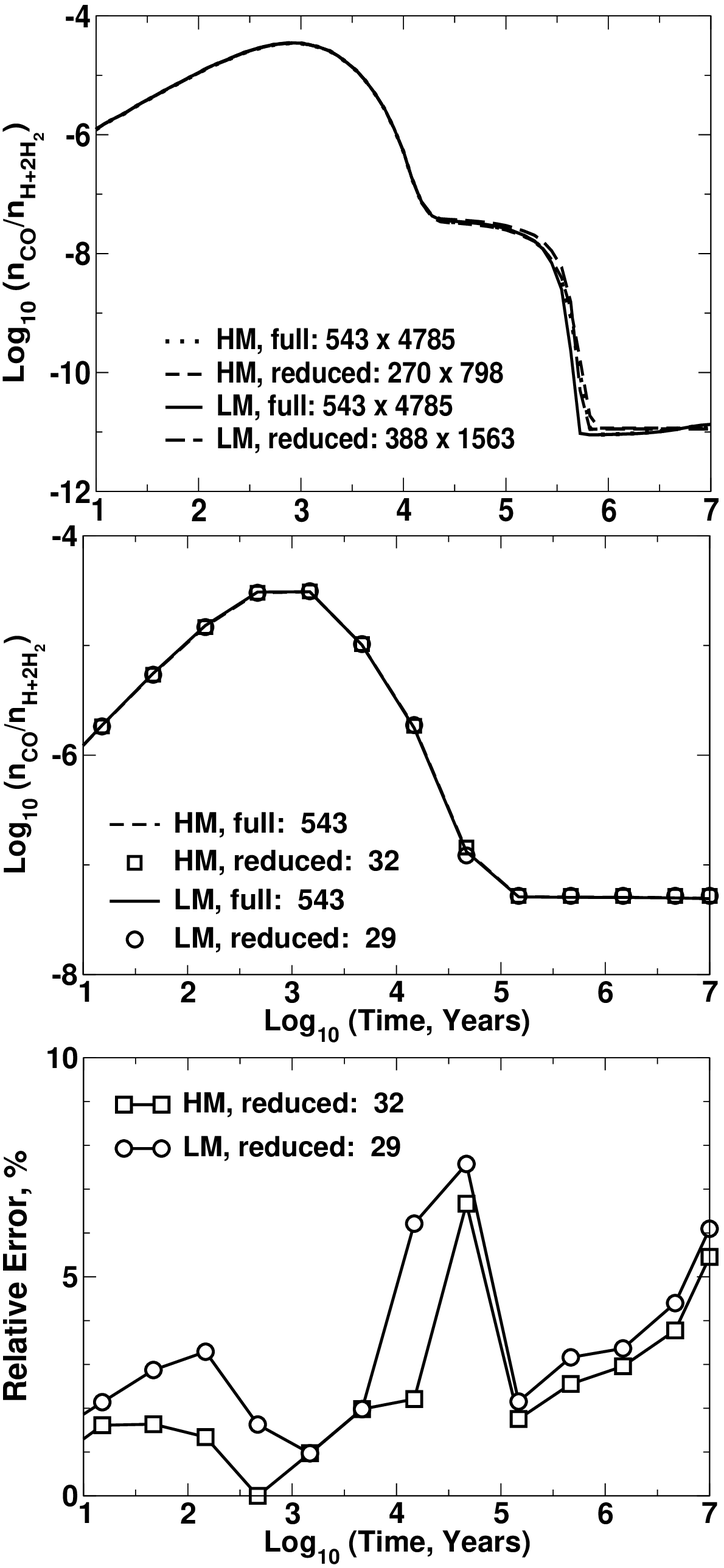}
\end{center}
\caption{Same as in Fig.~\ref{dens_dc_co} but for the DUST-DENS
({\em top} panel) and GAD-DENS ({\em middle} and {\em bottom} panels) models.}
\label{dens_dd_co}
\end{figure}

In the absence of surface reactions the evolution of carbon monoxide is 
simple (see Fig.~\ref{dens_dd_co}, middle panel). First, its abundance grows
steadily, at $t\la5\cdot10^3$~years reaching a value of $\sim 5\cdot10^{-5}$. 
Then the gas-grain interaction becomes important, and the CO molecules heavily 
deplete from the gas phase, forming significant mantles. The chemical equilibrium 
is reached at $10^5$~years with the gas-phase CO abundance of a few times $10^{-8}$.

The species-based reduction technique isolates two sets to be used in the high-
and low-metallicity cases of the GAD-DENS model. But in fact these sets are very similar
to each other, and one can use the same set containing about 30 species with high- and 
low- initial metallicities both in pure gas-phase models and with accretion/desorption
taken into account. The resulting uncertainties are smaller than $10\%$ for both 
metallicity cases (see Fig.~\ref{dens_dd_co}, bottom panel). 
Again, the computational time which is necessary to compute the evolution of 
CO is negligibly small.

The evolution of carbon monoxide in the situation, when surface reactions 
are considered, is shown in Fig.~\ref{dens_dd_co}, top panel. The gas-phase
CO abundance increases at early times and reaches the value of $\sim 10^{-4}$
at $t\sim10^3$~years. Then, the depletion of CO becomes efficient. At
$t\sim 10^4-10^6$~years the gas-phase CO abundance is determined by the balance
between accretion and desorption processes. Its value is almost the same as
the equilibrium CO abundance in the GAD-DENS model, i.e., a few times $10^{-8}$.
After $10^6$~years surface re-processing of CO takes over. It is converted to
other species, mainly, CO$_2$. At this stage the accretion of CO is no more
balanced by its desorption, and after some time the new, much lower
equilibrium value $\sim10^{-11}$ is established. Note that this behaviour is at least
partially caused by our simplified treatment of the gas-dust interactions. We adopt
the Hasegawa \& Herbst (\cite{HH93}) value for CO binding energy which is estimated
for a SiO$_2$ surface. In reality, after the CO mantle is formed, the binding energy of 
CO and other molecules may become smaller, so that desorption is still competitive with 
surface re-processing.

With the reaction-based reduction, we are able to select sets of 388 species and 1563 
reactions for the high-metallicity case and 270 species and 798 reactions for the 
low-metallicity case. The relevant computational gains are only about 3 and 16, 
respectively (see Table~\ref{dens_co2}). The error of the predicted CO abundances is 
less than $10\%$ during the entire evolutionary time. The number of species which are 
needed to estimate the CO abundance accurately in the high-metallicity case is again 
$\sim 50\%$ larger compared to the low-metallicity case. Reduced networks with smaller 
number of species are not usable in the calculations in this case. The reason is that
the relevant ODE systems are so stiff for the DVODE solver that the computational times 
are much larger compared to the case of the full chemical network.

\begin{table}
\caption{Reductions made for CO in a dense cloud with gas-grain chemistry 
included.}
\label{dens_co2}
\begin{tabular}{lllc}
\hline
 Model  & Important & Reduced & Speed \\
        & species   & network & gain \\
\hline
\multicolumn{4}{c}{Species-based reduction}\\
\hline
HM-GAD-DENS & CO            & $31\times148$     & $>100$ \\
LM-GAD-DENS & CO            & $29\times133$     & $>100$ \\
\hline
\multicolumn{4}{c}{Reaction-based reduction}\\
\hline
HM-DUST-DENS & CO & $388\times1563$ & 3 \\
LM-DUST-DENS & CO & $270\times798$ & 16 \\ 
\hline 
\end{tabular}
\end{table}

Concluding this section, we emphasise that if surface reactions are
considered in a chemical network, then its reduction for carbon monoxide is of
no practical use as the relevant computational speed gains are too small. 
However, if accretion and desorption of species are taken into account as the 
only gas-grain processes, then it is possible to find a few tens in half
a thousand species which must be kept in a reduced network in order to follow 
the evolution of CO without significant ($\ga 50-100\%$) errors.

\section{Discussion}

The purpose of the present paper is to check whether a typical astrochemical database 
contains redundant information that can be ignored in certain calculations regarding 
interstellar chemistry. The primary incentive for our work is the long time usually
needed to complete a more or less detailed model of interstellar chemistry in
dynamically evolving media. In many cases the evolution of only a few
or even one species are of the main interest for an astronomer. The notable example
is the degree of ionisation in dense interstellar clouds which is regulated by the
abundances of a few dominant ions. It has been shown by many authors that in this case
the relevant chemistry is governed by a rather restricted set of chemical processes
(e.g. Oppenheimer \& Dalgarno~\cite{od1974}; Ciolek \& Mouschovias~\cite{cm1995}).

Of course, one needs a mathematically rigorous way to ``extract''
only those species and reactions from the entire network that are needed
to reproduce the evolution of a species under investigation within a reasonable numerical
accuracy (say, $<100\%$). Such a reduction would decrease the number of species in the 
network and, hence, the relevant computational time. Using the DVODE solver, we found that
this time is proportional to roughly the third power of the number of the involved 
chemical species. Thus, by decreasing the number of species in the model by just a factor 
of two, one can obtain an order of magnitude computational speed gain! Besides, the 
reduction would help to understand the role of a particular chemical compound or reaction
for the evolution of an important species. However, one must bear in mind that the cost of
the reduction is that the relevant reduced network has a certain range of applicability, 
i.e. it gives reliable estimates on the abundances of an important species for a specific 
range of the gas temperature, density, etc.

The first mathematically proven reduction of astrochemical networks is
made by RRPHH. Their intention was to isolate a set of species and reactions which are
needed to follow the evolution of the CO abundance within the 30\% accuracy.
They considered physical conditions, typical of translucent regions, i.e.,
$n_{\mathrm H}=500-10^5$~cm$^{-3}$, $T=10$~K, $A_{\mathrm V}=0.5-5$ mag, with
``low metal'' initial abundances. They discriminated 69 necessary species participating
in 241 reactions from the Ohio State New Standard Model, including reactions of 
dissociative recombination on dust grains. They found that it is possible to decrease
further the number of necessary species and reactions to 33 and 116, respectively, by
the artificial modification of the network. It consisted in exclusion of all reactions
involving carbon chains longer than C$_2$H$_2^+$ and in introduction instead of few 
synthetic reactions to correctly predict the abundances of shorter carbon chains.

The same method was applied by RBHPR in order to identify reduced networks that
predict the degree of ionisation in interstellar clouds with less than 30\% uncertainty
under a wide range of physical conditions. They reduced two networks, namely, a ``full''
network of 218 species and 2747 reactions and a ``large'' network of 79 species and
1026 reactions. In the ``full'' network metals beside sodium were excluded, whereas the
``large'' network in addition excluded nitrogen and most species having more than two
nuclei heavier than hydrogen. High and low metal initial abundances were used.
They considered the following physical conditions: $T=20$~K,
$n_{\mathrm H}=500-10^5$~cm$^{-3}$, $A_{\mathrm V}=1-10$ mag. The small combined reduced 
network having 28 species and 71 reactions was found for the case of low metallicity. In 
the high metallicity case similar network of 29 species and 65 reactions was isolated with 
the artificially simplified sulphur chemistry.

We analysed the possibility of reducing the number of species and reactions in the
UMIST\,95 database by two different methods that we refer to as species-based and 
reaction-based techniques. Our primary goal is to compare the efficiency of these 
techniques under physical conditions typical of dark regions. We study the applicability 
of both methods in the case, when the complicated gas-grain interactions and the surface 
reactions are taken into account. Having in mind possible applications, we tried to make 
these methods as automatic as possible. Unlike RRPHH and RBHPR, we did not apply 
artificial modifications to the reduced networks. While such modifications can be useful 
in certain cases, usually it is not a trivial task and can be fraught with difficulties 
when a wide range of physical conditions is considered. Thus, we left the reduction 
entirely to the reduction techniques.

Both techniques proved to be efficient enough to reduce significantly gas-phase
chemical networks needed to compute the ionisation degree or CO abundance in diffuse and
dense molecular clouds with a reasonable accuracy of $<30\%$. If the accretion-desorption 
processes and surface reactions are accounted for, the reduction becomes less effective 
since the relevant chemistry is too complicated. Even though the number of chemical 
reactions in the reduced networks could be decreased by a few times, making their analysis
easier, the computational speed gains are too small in these models to make this reduction
of particular interest for chemo-dynamical modelling. Though, at least for the 
ionisation degree, the situation may be alleviated by the fact that at high densities when
depletion is effective, grain charging processes are more important for the fractional 
ionisation than the chemistry.

We found that the reaction-based technique is more effective compared to the species-based
technique (see Tables~\ref{diff_e}--\ref{gad_e}). The number of reactions is at least 
three times smaller for the reaction-based reduction than for the species-based reduction
which is natural to expect as reaction rates are not discriminated in the latter method.
But the number of species is also smaller for the reaction-based technique in most models.
When the gas-dust interaction is taken into account, the reaction-based technique is even
more effective than the species-based technique, as it provides the simultaneous reduction
of the number of species and reactions and proper treatment of the gas-grain processes.
Unfortunately, it is impossible to compare our results for the ionisation degree directly 
with the results of RBHPR. One reason is that they considered moderate values of visual 
extinction $A_{\mathrm V}=1-10$ mag, typical of translucent clouds, whereas we dealt with 
obscured regions, where $A_{\mathrm V}>10$ mag. Another issue is that RBHPR started the
reduction from a somewhat smaller ``full'' network of 218 species and 2747 reactions,
while we considered the network of 395 species involved in 3864 reactions. Nonetheless,
the resulting reduced networks have comparable sizes, namely, 29 species and 65 reactions 
(RBHPR, see Table~4) and 58 species and 111 reactions (this work, Table~3) in 
the case of high metals. In the low metallicity case the reduced network by
RBHPR has 28 species involved in 71 reactions, compared to our network
of 73 species and 169 reactions.

In the case of reduction made for CO, the situation is a bit more diverse.
In a diffuse cloud with pure gas-phase chemistry, the species-based 
reduction technique overcomes the reaction-based method, but at the expense of higher 
uncertainties (see Table~\ref{diff_co} and Fig.~\ref{diff_dc_co}). In a dense cloud
with pure gas-phase chemistry, the reaction-based technique allows us to extract
the reduced network which is very small (Table~\ref{CO}) but still functional under 
a very wide range of physical conditions. This network is smaller than the one found by 
RRPHH, though it has been extracted from the larger full network. Also, it has a wider 
range of applicability.

The next question to ask is how to apply these methods in dynamical calculations.
One can envisage two different strategies to pursue this goal. One way is to reduce the 
number of species and reactions at the outset and then to use the reduced network during 
the entire computational time. Another strategy would be to perform the network analysis 
in the ``real-time'' mode, during the dynamical evolution.

The first strategy is obviously more cost-effective as it does not add any overhead to the
computational time. The reduced network must be prepared in advance and checked against 
the range of physical conditions that are expected in the dynamical model. Of course, this
check is only possible under simplified conditions (otherwise it makes no sense). For 
example, if the range of densities (temperatures) is expected in the model, one may check 
that the reduced network is valid at its upper and lower boundaries, hoping that it is 
valid as well for any densities (temperatures) in between. Of our two techniques, the 
reaction-based approach works in this mode, as to construct the reduced network the 
abundances during the entire evolutionary time are analysed. The resultant network 
accounts for both the early and later chemistry.

The second method seems to be more robust. After the ability of the reduction technique to
remove the excessive information from the chemical network is checked, one can perform the
reduction, for example, at each dynamical time step or even less often. The full chemical 
network must be taken into account every time, with current abundances, temperature, 
density, etc. One may want to identify not only species that can be ignored but species 
with quasi-stationary abundances as well. Our species-based reduction technique is well 
suited to be used at a given $t$ with the current values of parameters because, as 
we mentioned above, the sharp jump in $B_i$ values allow to distinguish necessary and 
unnecessary species clearly. Usually, there is no clear boundary between needed and 
unneeded reactions (see Fig.~2), so one has to readjust cut-off threshold several times 
before optimal reduction is found. Thus, in some cases, the reaction-based method is more 
time-consuming than the species-based approach which is critical if the method is to be 
invoked from the dynamical code more than once.

Neither method guarantees that the same accuracy, as found in test cases, is preserved
during the real computations. With the first method one cannot be sure that conditions, 
under which the reduced network is not valid, are never met in the dynamical model. The 
second method properly takes into account the current conditions but is more sensitive to 
the history of abundance changes. Consider the ionisation degree in the dense cloud when 
freeze-out is taken into account (high-metallicity case). One of the dominant ions at later
times is N$_2$H$^+$. As a necessary species, it is included in the network at the time 
when metal depletion becomes important (late stages). However, the abundance of this ion 
cannot be followed with any accuracy as it is absent in early-time reduced networks,
and its entire previous evolution is lost. 
So, both strategies must be applied with care, taking into account possible uncertainties.

We note in passing that the decreased accuracy of the reduced networks may not be such an
issue. Many reactions in the UMIST\,95 have poorly measured or estimated rates,
with typical uncertainties $\la 25-100\%$ or more (Millar et al.~\cite{UMIST95}). This 
introduces unknown global errors in the modelled species abundances. On the other hand, 
modern observations usually provide column densities of abundant species with 
uncertainties of few tens of percents. Therefore, at nearly all astrophysically important 
situations accuracy of $\sim50\%$ seems to be quite appropriate.

\section{Conclusions}

We developed a robust method to reduce simultaneously the number of species
and reactions in chemical networks. The straightforward application for such
reduced chemical networks would be the modelling of the evolution of magnetised
protostellar clouds or protoplanetary discs, when it is necessary to compute the
fractional ionisation self-consistently with dynamical processes.

Investigating the applicability of this reduction approach, we considered
the conditions of diffuse and dense molecular clouds and utilised the UMIST\,95 database of
chemical reaction rates. For the sake of comparison, we performed the same reduction
with the objective reduction technique. The new reaction-based way of reduction
proved to be more efficient and accurate but more time-consuming as well.

It is found that the number of species and reactions that
are needed to follow the evolution of the CO abundance or the ionisation degree can be
greatly decreased in the case of pure gas-phase chemistry, implying
computational speed gains of more than a few hundred. For instance, to follow accurately 
(with $<15\%$ uncertainty) the evolution of the CO abundance in a dense cloud, it is
enough to retain only 8 species involved in 9 reactions in the chemical network.
However, as soon as the gas-grain interactions and surface
chemistry are taken into account, reduction is modest and corresponding
acceleration factors are around 10.

\begin{acknowledgements}
  DS was supported by the German 
  \emph{Deut\-sche For\-schungs\-ge\-mein\-schaft, DFG\/} project 
  ``Research Group Laboratory Astrophysics'' (He 1935/17-1), the work 
  of DW was supported by the INTAS grant YS 2001-1/91 and the RFBR 
  grant 01-02-16206. We thank the anonymous referee for 
comments and suggestions.
\end{acknowledgements}
\appendix
\section{Adopted input parameters}
We use the standard DVODE package designed to solve systems of
ordinary differential equations. The relevant input parameters adopted in 
our computations are the following:
\begin{itemize}
\item absolute error $\epsilon_\mathrm{abs}=2.20\cdot10^{-16}$;
\item relative error $\epsilon_\mathrm{rel}=10^{-8}$; 
\item Jacobian computed by the solver (internally).
\end{itemize}
We found that the relative error $\epsilon_{rel}=10^{-8}$
is enough to give reliable estimations of the abundances of the species. 

We use the g77 compiler and SuSe Linux 7.3 installed on the Pentium~IV
2~GHz, 1~GB RAM PC as well as Compaq Visual Fortran 6.5 compiler on
Pentium~III, 850~MHz, 256~MB RAM PC. The typical time necessary to run a 
model is only a few minutes in the case of gas-phase chemistry and can 
reach half an hour if the gas-grain interactions are taken into account.

\end{document}